\documentclass[a4paper]{article}

\usepackage{icrc2013}
\usepackage[english]{babel} 

\title{Open-structure Composite Mirrors for the Cherenkov Telescope Array}

\shorttitle{Open-structure Composite Mirrors for CTA}

\authors{
Michal Dyrda$^{1}$,
Jerzy Michalowski$^{1}$,
Jacek Niemiec$^{1}$,
Marek Stodulski$^{1}$
for the Cherenkov Telescope Array Consortium.
}

\afiliations{
$^1$ Institute of Nuclear Physics, PAS, ul. Radzikowskiego 152, 31-342 Krak\'ow, Poland \\
}

\email{Michal.Dyrda@ifj.edu.pl}

\abstract{The Cherenkov Telescope Array (CTA) Observatory for high-energy gamma-ray astronomy will comprise several tens of imaging atmospheric Cherenkov telescopes (IACTs) of different size with a total reflective area of about 10,000 m$^2$. Here we present a new technology for the production of IACT mirrors that has been developed in the Institute of Nuclear Physics PAS in Krakow, Poland. An open-structure composite mirror consists of a rigid flat sandwich support structure and cast-in-mould spherical epoxy resin layer. To this layer a thin glass sheet complete with optical coating is cold-slumped to provide the spherical reflective layer of the mirror. The main components of the sandwich support structure are two flat float glass panels inter spaced with V-shape aluminum spacers of equal length. The sandwich support structure is open, thus enabling good cooling and ventilation of the mirror. A special arrangement of the aluminum spacers also prohibits water being trapped inside. The open-structure technology thus represents a novel cost-efficient approach that does not require the perfect sealing needed in closed-type mirrors. In addition, the technology enables the application of a dielectric coating. Full-size prototype mirrors designed for the medium-size CTA telescope will be presented together with the results of recent optical tests.}


\begin{document}
\maketitle

%Begin a section.
\section{Introduction}

Current imaging atmospheric Cherenkov telescopes (IACTs): H.E.S.S., MAGIC and VERITAS, use reflective dishes, which are segmented into mirror facets. The performance of these telescopes is highly dependent on the effective mirror area and the quality of the Cherenkov light shower images, the later is closely related to the size of the point source image. 

One of the examples of mirror technology for Cherenkov telescopes is a cold-slumped glass mirror coated with an aluminum. These kinds of mirrors are used on H.E.S.S. and VERITAS telescopes. The main challenge of this technology is a degradation of the reflective layer exposed to environmental conditions, which cause a need for re-coating after some time \cite{bib:foerster1}. MAGIC telescope use different mirrors, namely diamond-milled aluminum facets with quartz coating. The main issue with this technology is that production of such mirrors is time consuming and expensive. One should keep in mind that the available mirror production technologies may not be sufficient for the CTA observatory, where several tens of different type telescopes will be built. All possible mirror technologies developed specially for the CTA telescopes are described in \cite{bib:foerster}.

Technology for the production of mirrors for CTA telescopes has been developed at INP PAS for the last four years. The manufactured mirror prototypes are designed for medium size telescopes (MST), which is a classical Davies-Cotton construction \cite{bib:davies}. With the focal length of 16.08 m, the MST mirrors should have a total reflectivity greater than 85\% in the wavelength range between 300 and 600 nm. The pixel size for the MST telescope (photomultiplier plus light cone) is 50 mm and the requirement for the MST mirror facets is that more than 80\% of the reflected light should be focused within 1/3 of the pixel size (d80 - 80\% containment radius), is $\sim$ 17 mm.

\begin{figure*}[!t]
\centering
\includegraphics[width=\textwidth]{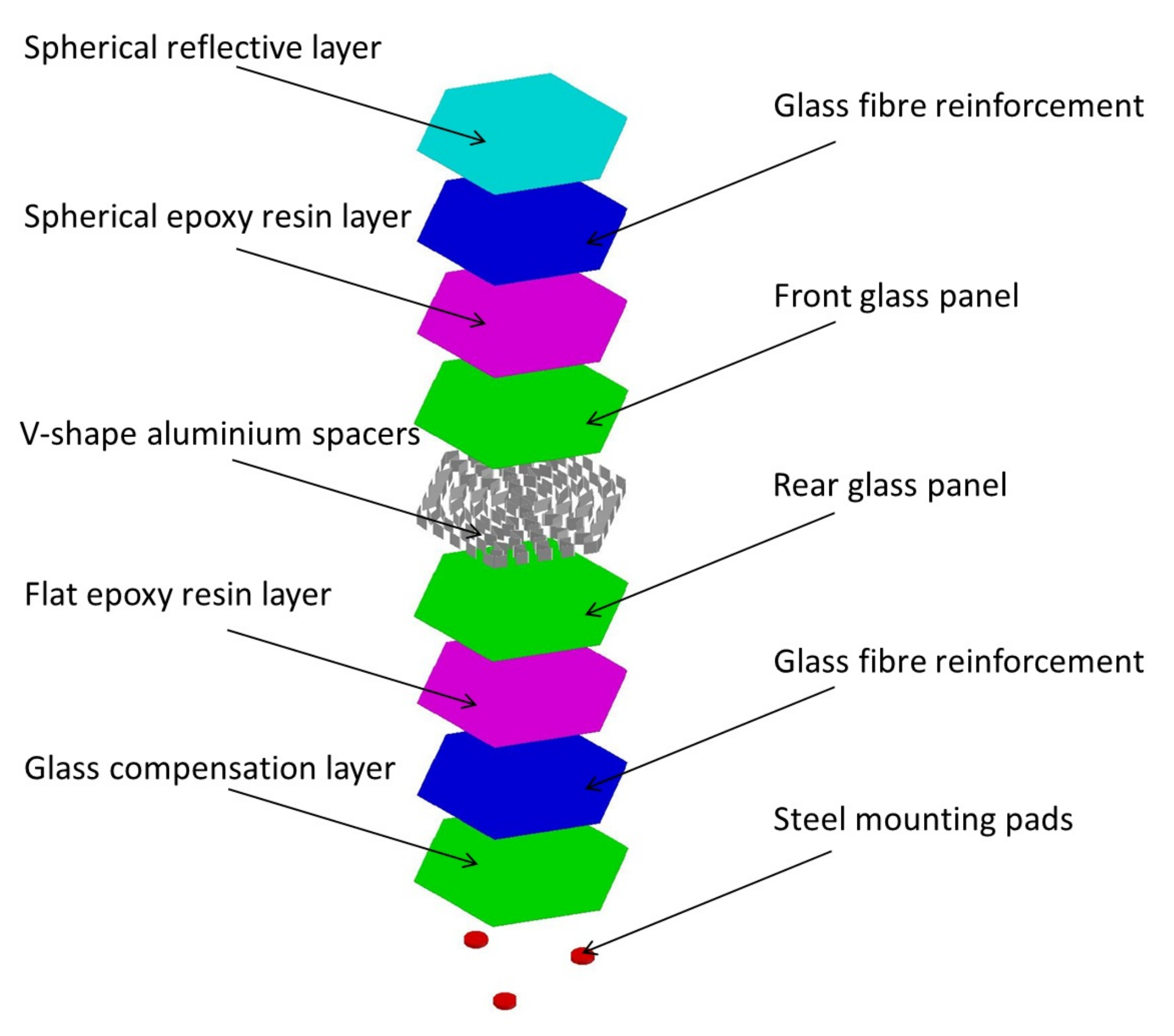}
\caption{Open-structure composite mirror designed at INP PAS, Krakow.}
\label{fig1}
\end{figure*}

\section{Technology Description and Design Status}

\begin{figure*}[!t]
\centering
\includegraphics[width=0.4\textwidth]{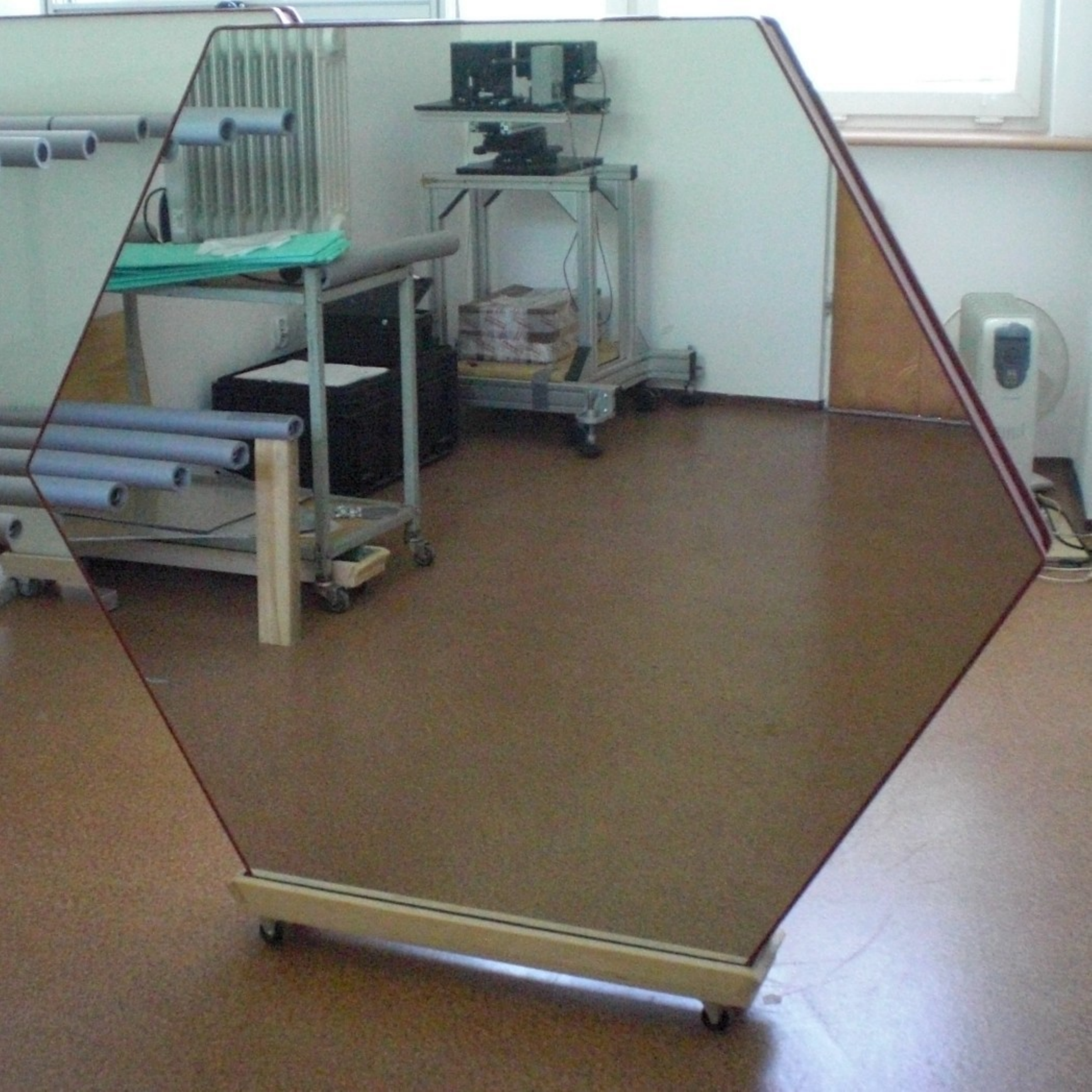}\includegraphics[width=0.4\textwidth]{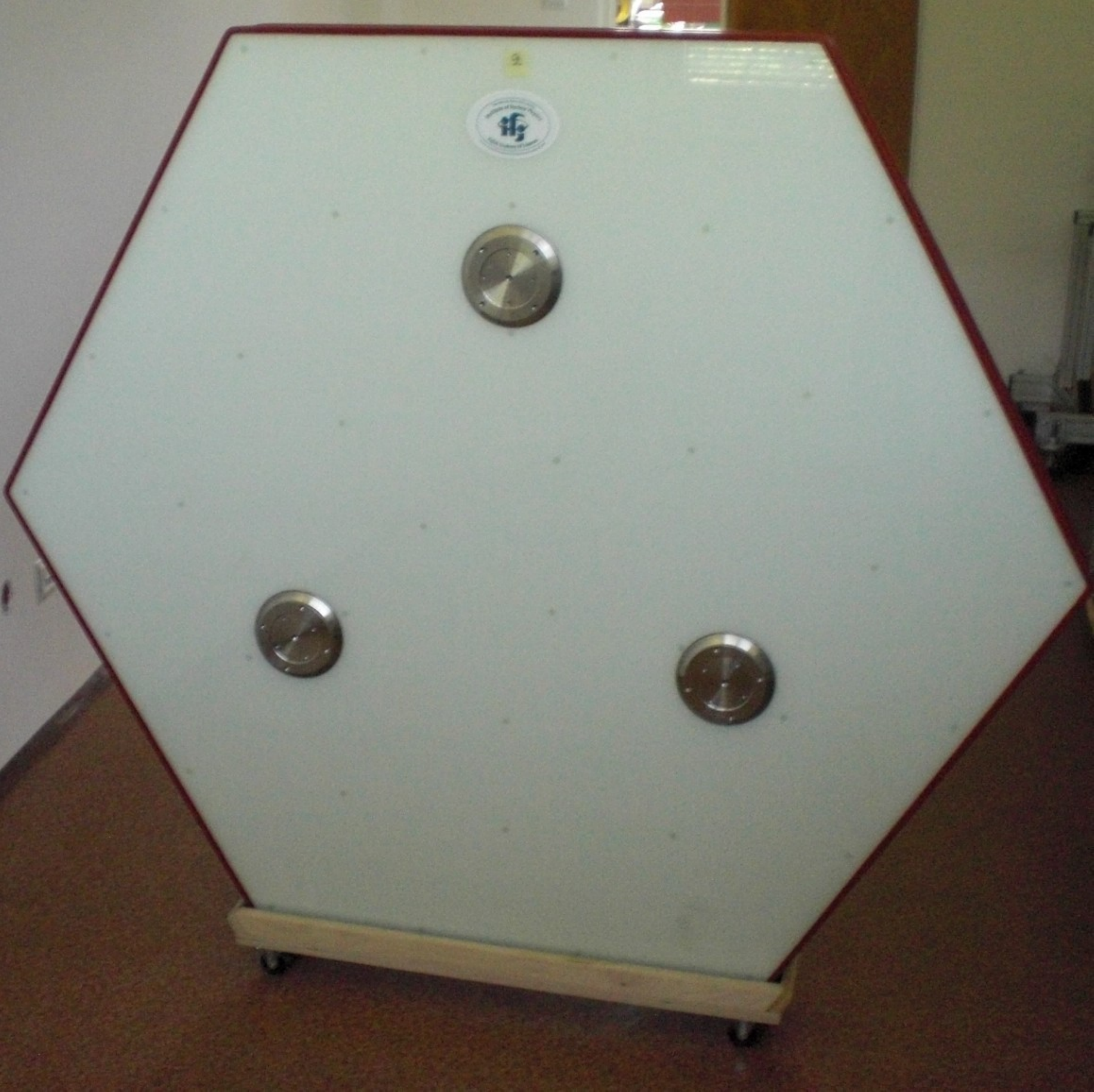}
\includegraphics[width=0.4\textwidth]{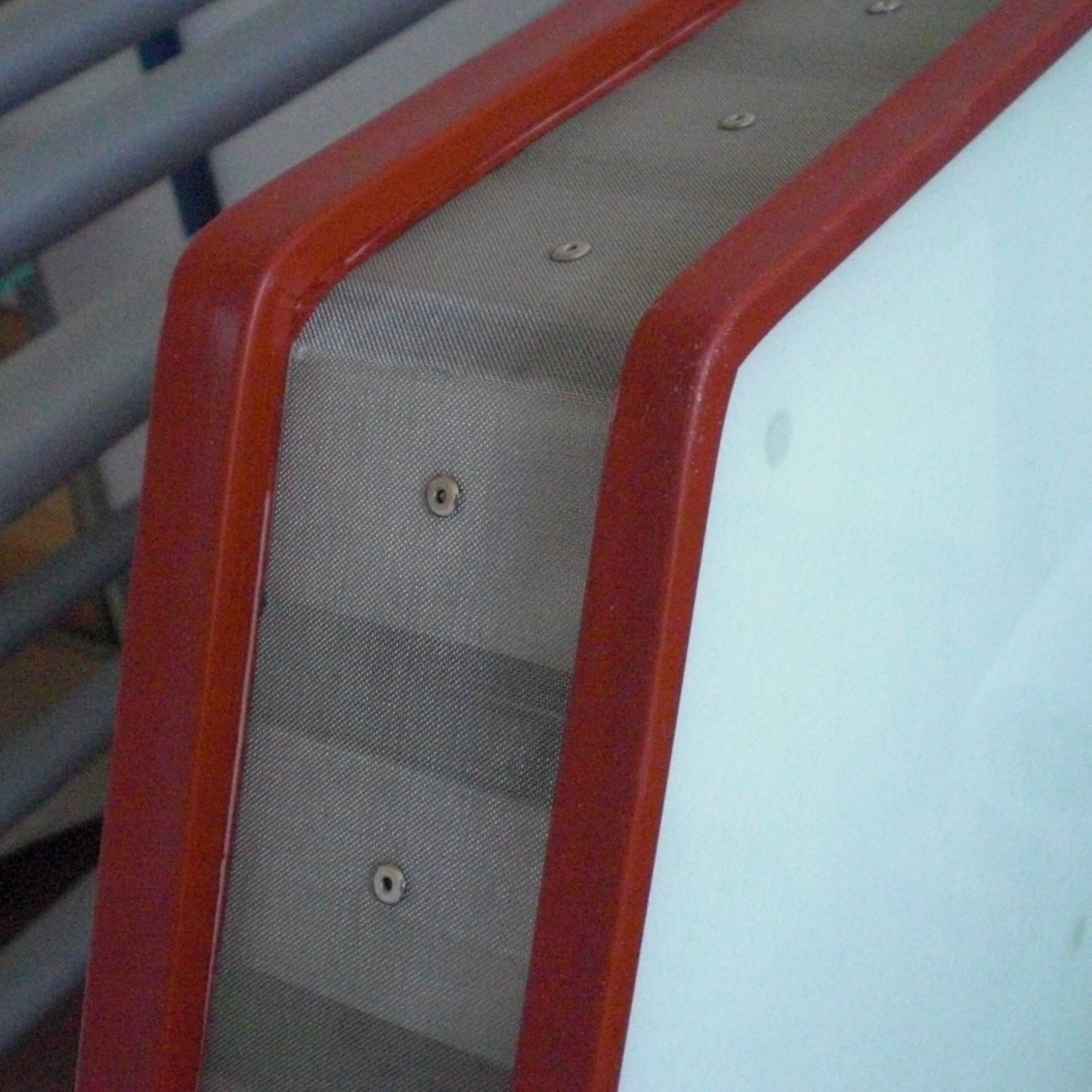}
\caption{The front of an open structure mirror with reflective layer (top left) and the rear side with mounting pads (top right). Visible are the protective silicon rubber and stainless steel mesh (bottom panel).}
\label{fig2}
\end{figure*}

The open-structure mirror consists of a sandwich support structure, a spherical resin layer and a spherical glass reflecting layer, as shown in figure \ref{fig1}. The sandwich support structure consists of two flat glass panels separated by the spacers, which could be aluminum profiles of L, C, S, V or Z shape and wall thickness between 0.5mm and 1.0 mm. However, to prevent water becoming stuck inside the sandwich structure, V-shape aluminum profiles were selected. The profiles are glued to the two flat panels with epoxy resin. The mirror is hexagonal in shape with size 1.2 m flat-to-flat. Both panels are made of ordinary float glass. The front and the rear panel thickness is 2 mm. The spacers have wall thickness of 0.8 mm and length of 70 mm. Each arm of the V-shape is 40 mm in length and the angle between them is 60 degrees. The total number of aluminum profiles for each mirror is 470. The use of a surface table during the gluing process yields a very flat sandwich structure, whose thickness diverge by no more than $\pm$ 0.1 mm. 

Another compensation glass sheet of thickness 2 mm with its resin layer, of size 3 mm, is glued to the rear panel to ensure the geometrical stability and stiffness of the structure during thermal cycles \cite{bib:chadwick}. Two fibreglass tissues are used to reinforce the substrate structure and to improve the resistance to the mechanical impact. The first fibreglass tissue is in between the spherical reflective layer and the spherical epoxy layer. The second tissue is in between the flat layer of epoxy resin and the glass compensation layer (See in figure \ref{fig1}). Three stainless steel pads are glued to the compensation glass layer. These three points mounting interface system is 320 mm from the mirror centre to enable mounting of the actuators designed for CTA mirrors, and is shown in figure \ref{fig2}. The use of the V-shape spacers in the sandwich structure provides good ventilation and cooling of the mirror glass panels. To prohibit water overwhelming inside of the mirror, the V-shape spacers are organized such that the water flows almost freely through the structure. This is a novel solution different from the commonly applied closed aluminum honeycomb substrates. The latter require a perfect sealing of a mirror with additional sidewalls so that water cannot penetrate inside the honeycomb and possibly cause damages to the structure. The open-structure mirror has an axial symmetry and has to be mounted properly on the telescope dish support structure. Markers indicating the correct mirror orientation are used to ensure appropriate installation. Stainless steel mesh is attached to the sidewalls to protect the sandwich structure against contamination by insects or bird waste.

The open-structure mirrors are to be used on the medium size Davies Cotton telescope for CTA and hence their radius of curvature should be 32.16 m \cite{bib:baehr}. To ensure the convex surface of the mirror a layer made of compound of the epoxy resin and fillers is cast onto a front panel in a high precision mould. This mould, diameter of 1.4 m, is specially designed for this purpose. It is equipped with vacuum and heating systems, and is mounted on the steel support. The spherical resin layer, after hardening at room temperature, allows the reproduction of the exact shape of the master mould. A final open-structure prototype mirror support structure is depicted in figure \ref{fig2}. After the spherical resin layer has been cast, a reflective layer made of Borofloat 33 glass sheet \cite{bib:schott}, 2 mm thick, is cold-slumped to it in the same mould. The Borofloat 33 glass reflective layer was coated with Al+SiO$_2$+HfO$_2$+SiO$_2$ by the German company BTE prior to gluing to the substrate. In the last step the special silicone rubber, resistant over a wide temperature range, from $-$60 to $+$260 [$^o$C], is glued to the mirror sidewalls to protect the mirror against damage during the transportation and mounting processes. The silicone rubber serves also to the front and rear panels to protect the prototype against the water penetration into the epoxy resin layer. The rubber sealing is shown in the bottom panel of figure \ref{fig2}. Technology developed at INP also enables us to use of a dielectric coating, which provides high durability of the reflective surface. One mirror with a dielectric reflective panel is under development at INP PAS.

The current technology enables to build a mirror of up to ~2 m in size up to $\sim$ 2 m and the only limitation is the availability of a high precision mould as well as the reflective panels such as Borofloat 33 glass or dielectric-coated ones within an affordable budget. The sequence of operations described above can be used to produce mirrors within wide range of convex surfaces, but one should bear in mind that the increase in the flat-to-flat size of the mirror results in the increase of the minimum curvature radius.

\section{Optical Test Results}

\begin{figure}[!h]
\centering
\includegraphics[trim= 1cm 9cm 1cm 8cm, clip=true, width=0.4\textwidth]{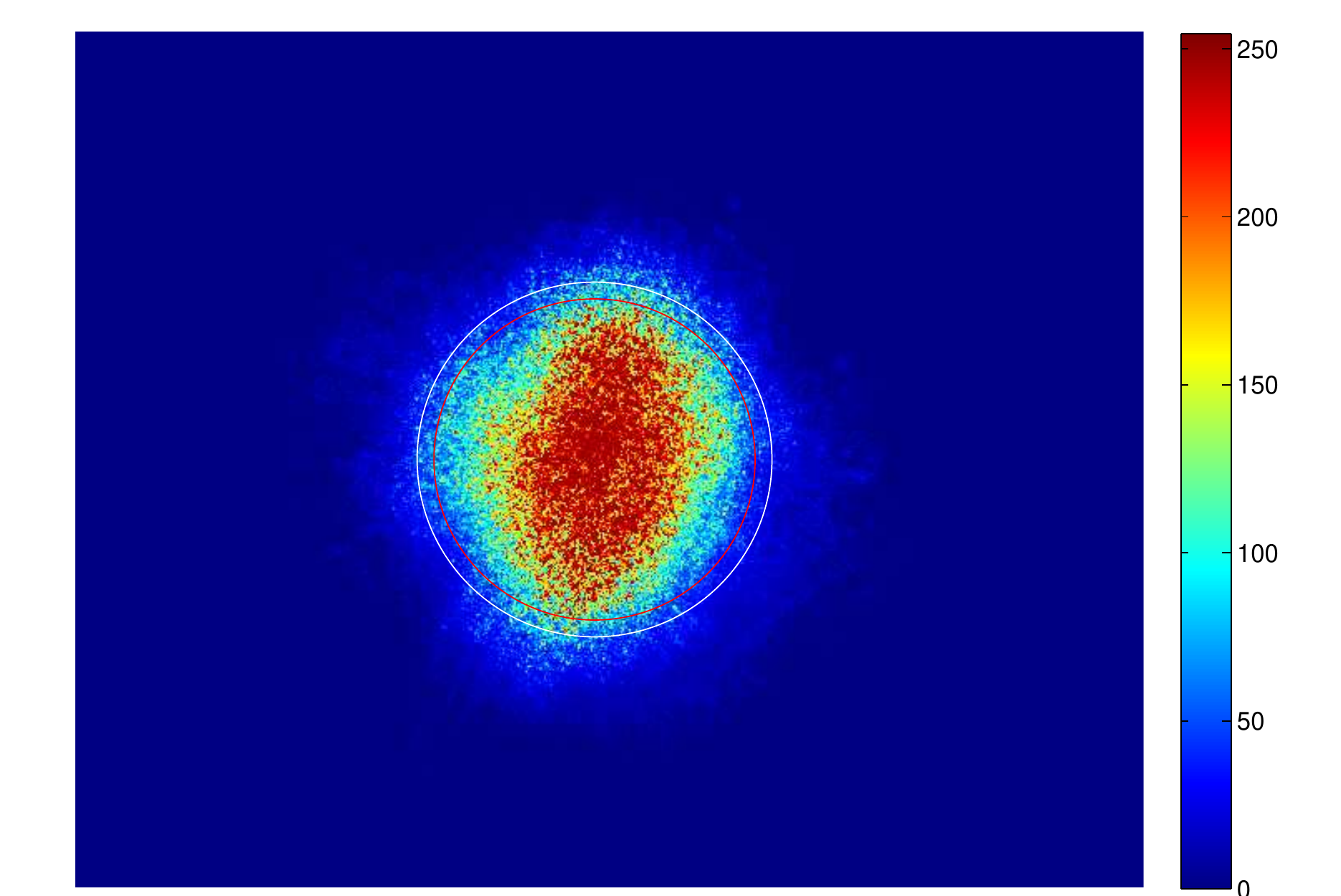}
\caption{PSF measurements for one mirror designed at INP PAS, Krakow, see text for details.}
\label{fig3}
\end{figure}

\begin{figure*}[!t]
\centering
\includegraphics[clip=true,width=0.4\textwidth]{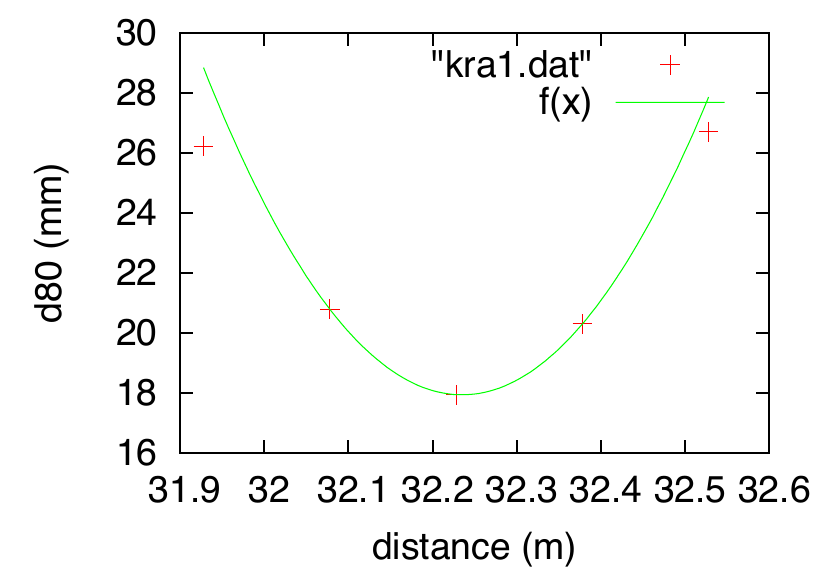}\includegraphics[clip=true,width=0.4\textwidth]{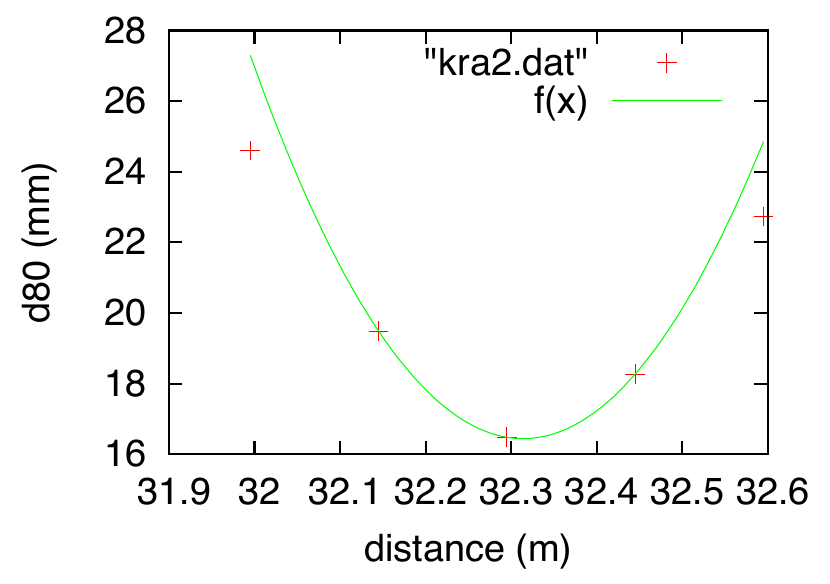}
\includegraphics[clip=true, width=0.4\textwidth]{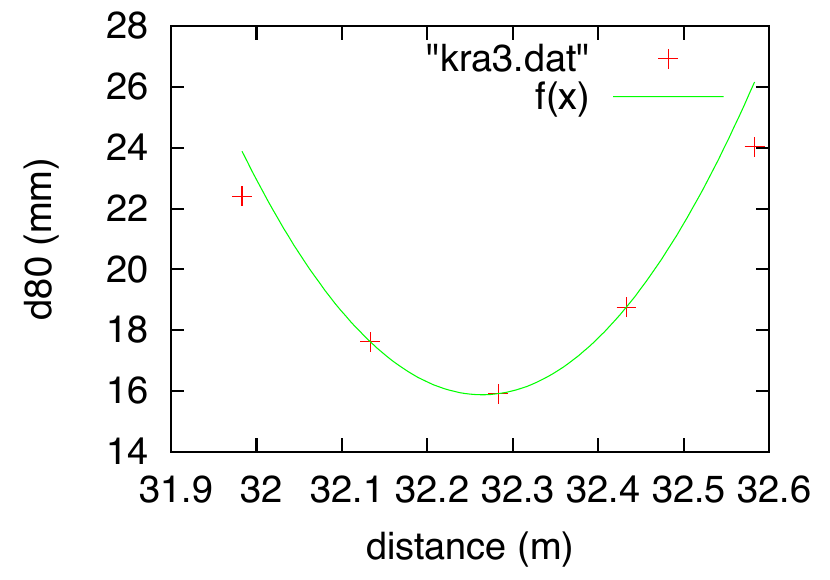} \includegraphics[clip=true, width=0.4\textwidth]{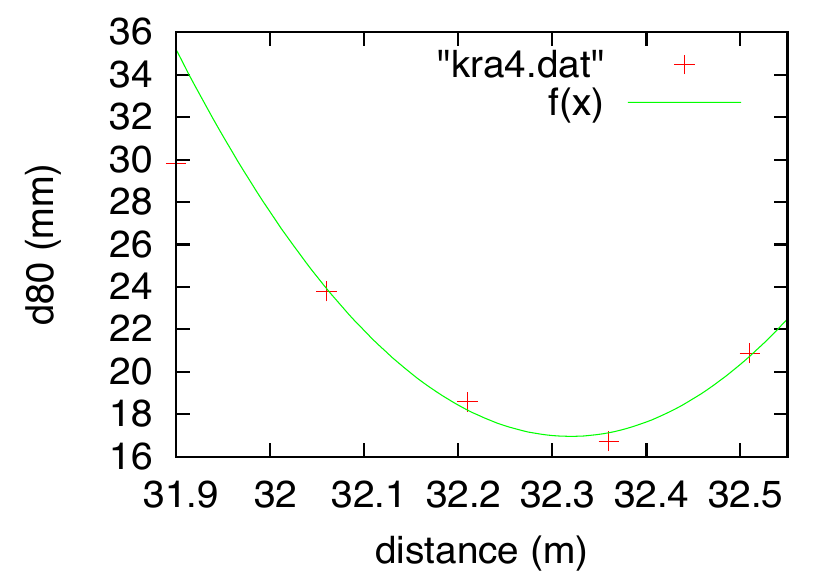}
\caption{Two focal length scan of four open structure mirror prototypes done. Red crosses denote measurements results and the fitted function is a green line. The inferred two focal length values are: 32.23 m, 32.31 m, 32.26 m and 33.33 m for each prototype, respectively.}
\label{fig4}
\end{figure*}

Nine hexagonal prototype mirrors were manufactured between March and May 2013. Preliminary tests were performed on all of them to measure the Point Spread Function (PSF). The measurements have been done with the test bench, which was designed and manufactured by the Space Research Centre PAS in Warsaw. The test bench consists of a blue laser, emitting at a wavelength of 405 nm, a specially designed jig to mount the mirrors for measurements and a CCD camera with software for image capture and processing. The mirrors are placed at the distance equals to two nominal focal lengths (32.16 m) and the preliminary PSF measurement is done. The focal length of a mirror can be determined from its PSF measurement, since at focal length the mirror PSF will be at its minimum.

An example result of PSF measurement is shown in figure \ref{fig3}, which was done for the 7th mirror prototype developed at INP PAS. The PSF spot (d80 - the radius of the circle in which 80\% of the reflected light energy is contained) is marked with a red circle, and for this particular mirror prototype d80 = 15.28 mm, which can compared with the CTA requirement for the MST mirrors, that d80 $<$ 17 mm. The two focal lengths distance measured for this sample is 32.25 m, compared with the nominal value of 32.16 m. The analysis of the rest of mirror prototypes gives a PSF value between 15.28 mm and 17.21 mm.

The first four open structure mirrors were sent to Erlangen Centre for Astroparticle Physics in April 2013 in order to perform independent optical tests. The results of the scans at twice the focal length are shown in figure \ref{fig4}. The nominal two focal length for those mirrors should be equal 32.16 m. Five measurements of the d80 were made of each prototype mirror in the vicinity of the nominal focal length, and a quadratic function was fitted to obtain the minimum value of the PSF and hence the value of twice the focal length. The inferred values of twice the focal length values vary form 32.23 m to 32.33 and are in good agreement with the nominal value of 32.16. The deviation from the nominal value arises because the radius of curvature of the master mould is a few cm bigger than the required value 32.16 m, which is a production defect. The measured d80 values for these four mirror prototypes, obtained at the inferred two focal-length distance, are between 16.12 mm and 18.46, which can be compared with the results obtain at INP PAS. 

\section{Conclusions}

A novel mirror technology, for Cherenkov telescopes has been proposed, which links the standard cold-slump approach with modern compound. The manufacturing steps developed are independent of the coating processes and hence different reflective layers can be used. The other important advantage of the mirror technology presented in this paper is its open architecture, which does not face the well-known problems of other closed structures and honeycomb technologies. The open structure of the mirrors make them naturally pressure-equalize, when placed at high altitude in comparison with closed architectures. The open-structure mirrors samples underwent some outdoor tests during the wintertime in the UK. These results are described in \cite{bib:chadwick}. 

Nine mirror prototypes have been produced at INP PAS. Preliminary optical tests show a very good agreement with CTA specifications. The first four prototypes are under tests in Erlangen and the second four were sent to DESY Institute in Zeuthen to be mounted on the first MST prototype telescope. The last, 9th, mirror will be send to Argentina for one year of outdoor tests.

\vspace*{0.5cm}
\footnotesize{{\bf Acknowledgement:}{ This work was supported by The National Science Centre through project DEC-2011/01/M/ST9/0189. We gratefully acknowledge support from the agencies and organizations listed in this page: http://www.cta-observatory.org/?q=node/22}}

\end{document}